\documentclass[conference]{IEEEtran}
\usepackage{amsmath,amssymb,threeparttable,placeins}
\usepackage[ruled,vlined]{algorithm2e}
\usepackage{lipsum}
\usepackage{bm}
\usepackage{multicol}
\usepackage{array}
\usepackage{algorithmic}

\usepackage[mathscr]{euscript}
\usepackage[dvips]{graphicx}
\usepackage[usenames]{color}
\usepackage{amsfonts}
\usepackage{latexsym}
\usepackage{subfigure}
\usepackage[format=hang]{subfig}
\usepackage[justification=centering]{caption}
\usepackage{floatrow}
\usepackage{multirow}
\usepackage{graphicx}
\usepackage{epstopdf}
\usepackage[english]{babel}
\DeclareMathAlphabet{\mathpzc}{OT1}{pzc}{m}{it}

\newtheorem{lemma}{{{\textit{Lemma}}}}

\newtheorem{definition}{\textit{Definition}}

\newtheorem{remark}{{{\textit{Remark}}}}

\newtheorem{example}{{{\textit{Example}}}}

\def\BibTeX{{\rm B\kern-.05em{\sc i\kern-.025em b}\kern-.08em
    T\kern-.1667em\lower.7ex\hbox{E}\kern-.125emX}}
\begin{document}

\title{Root Cross $Z$-Complementary Pairs with Large ZCZ Width\\
}

\author{\IEEEauthorblockN{ Shibsankar Das}
\IEEEauthorblockA{\textit{The Department of Electrical Engineering} \\
\textit{ Indian Institute of Technology Kanpur}\\
\textit{Uttar Pradesh, India}\\
Email: \textit{shibsankar@iitk.ac.in}}
\and
\IEEEauthorblockN{ Adrish Banerjee}
\IEEEauthorblockA{\textit{The Department of Electrical Engineering} \\
\textit{ Indian Institute of Technology Kanpur}\\
\textit{Uttar Pradesh, India}\\
Email: \textit{adrish@iitk.ac.in}}
\and
\IEEEauthorblockN{ Zilong Liu}
\IEEEauthorblockA{\textit{The School of CSEE} \\
\textit{University of Essex}\\
\textit{Colchester, United Kingdom}\\
Email: \textit{zilong.liu@essex.ac.uk}}
}

\maketitle

	\begin{abstract}
	In this paper, we present a new family of cross $Z$-complementary pairs (CZCPs) based on generalized Boolean functions and two roots of unity. Our key idea is to consider an arbitrary partition of the set $\{1,2,\cdots, n\}$ with two subsets corresponding to two given roots of unity for which two truncated sequences of new alphabet size determined by the two roots of unity are obtained. We show that these two truncated sequences form a new $q$-ary CZCP with flexible sequence length and large zero-correlation zone width. Furthermore, we derive an enumeration formula by considering the Stirling number of the second kind for the partitions and show that the number of constructed CZCPs increases significantly compared to the existing works.
\end{abstract}	



\section{ Introduction}
\IEEEPARstart{T}{he} concept of cross $Z$-complementary pairs (CZCPs) was first studied by Liu \textit{et al.} in \cite{2020Liu_CZCP}. By definition, every CZCP has 1) zero aperiodic auto-correlation sums for certain time-shifts around and away from the in-phase position and 2) zero aperiodic cross-correlation sums for certain time-shifts away from the in-phase position. Owing to their zero correlation zone (ZCZ) properties, CZCPs have been employed to design optimal training signals in spatial modulation (SM) system \cite{2011Renzo}, \cite{2015YangHanzo} to offer efficient channel estimation performance over frequency-selective channels \cite{2020Liu_CZCP}, \cite{2020LiuConfISIT}. 

In search of good training sequences for SM systems, Liu \textit{et al.} presented a direct construction for $q$-ary CZCPs of lengths $N=2^n$ based on generalized Boolean functions (GBFs) \cite{2020Liu_CZCP}. The constructed $q$-ary CZCPs from \cite{2020Liu_CZCP} are optimal as each has ZCZ width $Z = N/2$.  However, the sequence lengths of optimal binary CZCPs are restricted to the form of $2^{a+1}10^b26^c$, where $a, b,$ and $c$ are non-negative integers. Later, sub-optimal CZCPs with lengths of $10^b26^c$ are reported in \cite{2020CuilingAdhikaryCOML}. In \cite{2020AdhikaryTSP}, Adhikary \textit{et al.} proposed a direct construction based on GBFs for CZCPs with sequence lengths of the form $2^{n-1}+2$, where $n\geq 4$. Cross $Z$-complementary ratio $(\text{CZC}_{\text{ratio}})$, which refers to the ratio of the achievable ZCZ width and the theoretical maximum ZCZ width, was introduced in \cite{2020AdhikaryTSP} to  check the optimality of a given CZCP. Note that an optimal CZCP has $\text{CZC}_{\text{ratio}}=1$. Moreover, it was shown that the $\text{CZC}_{\text{ratio}}$ of the constructed $q$-ary CZCPs in \cite{2020AdhikaryTSP} is $1/2$. To design CZCPs with new sequence lengths and large $\text{CZC}_{\text{ratio}}$, Huang \textit{et al.} presented new binary CZCPs with sequence lengths of the form $2^{n-1}+2^{\nu+1}$ \cite{2021HuangCOML}, where $0\leq \nu\leq n-3$ and with  $\text{CZC}_{\text{ratio}}$ of $2/3$. In \cite{2021ZhangMesnager}, Zhang \textit{et al.} proposed two new approaches based on cascading sequences and BFs to construct binary CZCPs with varieties of $\text{CZC}_{\text{ratio}}$.  Recently, Yang \textit{et al.} proposed a construction \cite{2021Yang} for binary and quadriphase CZCPs of sequence lengths $2N$, with the condition that a binary length-$N$ (even)  $Z$-complementary pair exists. 

Motivated by the work of Huang \textit{et al.} \cite{2021HuangCOML}, we present a new construction method of  $q$-ary CZCPs with flexible sequence lengths and large $\text{CZC}_{\text{ratio}}$. Specifically,
\begin{itemize}
	\item We propose a systematic construction of new $q$-ary CZCPs with sequence lengths $2^{n-1}+2^{\nu+1}$ by using GBFs and two roots of unity, where $0\leq \nu\leq n-3$. The phases of the constructed CZCPs are determined by the given two roots of unity. Thus, the constructed CZCPs are called root CZCPs (RCZCPs).
	\item We derive the total number of constructed RCZCPs for a given sequence length. This shows that our design strategy has an advantage over the existing works with regard to the availability of larger class of RCZCPs.
	\item We also show that all the constructed $q$-ary RCZCPs have a large $\text{CZC}_{\text{ratio}}=2/3$. We point out  that our proposed construction method includes \cite[Th. 4]{2021HuangCOML} as a special case in the design of binary CZCPs. A promising research direction would be to show whether the proposed larger class of RCZCPs can be  used in multi-user SM systems to offer efficient channel estimation performance?
\end{itemize}

\section{Preliminaries}
\label{Pre:sec}
\subsection{Cross $Z$-Complementary Pairs (CZCPs)}
Given two $q$-ary sequences $\textbf{\textit{x}} = \big[x_0, x_1,\cdots ,x_{N-1}\big]$ and $\textbf{\textit{y}} = \big[y_0, y_1,\cdots ,y_{N-1}\big]$ of equal length $N$ over $\mathbb{Z}_{q}$, their aperiodic cross-correlation function (ACCF) at a time-shift $\tau$ is defined by
\begin{equation}\label{defi_ACCF}
\rho_{\textbf{\textit{x}},\textbf{\textit{y}}}(\tau):= \left \{
\begin{array}{cl}
\sum\limits_{k=0}^{N-1-\tau}\omega_q^{x_k-y_{k+\tau}};&~0\leq \tau \leq N-1;\\
\sum\limits_{k=0}^{N-1+\tau}\omega_q^{x_{k-\tau}-y_k};&~-(N-1)\leq \tau < 0;\\
0;& ~~\mid \tau \mid \geq N,
\end{array}
\right .
\end{equation}
where $\omega_q=e^{\frac{2\pi\sqrt{-1}}{q}}$ and $q\ (\geq2)$ is a positive integer. When the two sequences are identical, i.e., $\textbf{\textit{x}} = \textbf{\textit{y}}$, $\rho_{\textbf{\textit{x}},\textbf{\textit{y}}}(\tau)$ is called an aperiodic auto-correlation function (AACF) of $\textbf{\textit{x}}$.  For simplicity, the AACF of $\textbf{\textit{x}}$ will be  written as $\rho_{\textbf{\textit{x}}}(\tau)$.
\begin{definition}[\cite{2020Liu_CZCP}]
	\label{def:CZCP}
	Let $(\textbf{\textit{x}}, \textbf{\textit{y}})$ be a sequence pair with sequence length $N$. For an integer $Z$, let $\mathcal{T}_1=\{1,2,\cdots,Z\}$ and $\mathcal{T}_2= \{N-Z,N-Z+1,\cdots,N-1\}$. Then, $(\textbf{\textit{x}}, \textbf{\textit{y}})$ is said to be an $(N,Z)$-CZCP if it satisfies the symmetric zero (out-of-phase) AACF sums for the time-shifts over $\mathcal{T}_1\cup\mathcal{T}_2$ and zero ACCF sums for the time-shifts over $\mathcal{T}_2$. In short, an $(N,Z)$-CZCP satisfies the following two conditions.
	\begin{align}
	\label{cross:z:com:conditions}
	&C_1: \rho_{\textbf{\textit{x}}}(\tau)+\rho_{\textbf{\textit{y}}}(\tau)= 0,\ \forall \ |\tau| \in \mathcal{T}_1\cup\mathcal{T}_2; \nonumber \\
	&C_2:\rho_{\textbf{\textit{x}},\textbf{\textit{y}}}(\tau)+\rho_{\textbf{\textit{y}},\textbf{\textit{x}}}(\tau)= 0, \ \forall \ |\tau| \in \mathcal{T}_2.
	\end{align}
\end{definition}Throughout this paper, an $m\times n$ matrix with all zero entries is denoted by $\textbf{O}_{m\times n}$ and an $m\times n$ matrix with all one entries is denoted by $\textbf{1}_{m\times n}$.
\begin{definition}
	For a given $(N,Z)$-CZCP, the cross $Z$-complementary ratio $(\text{CZC}_{\text{ratio}})$ is defined by
	$\text{CZC}_{\text{ratio}}=Z/Z_{\text{max}}$, where $Z_{\text{max}}$ is the theoretical maximum value of $Z$.
\end{definition} Clearly, $\text{CZC}_{\text{ratio}}\leq 1$ as $Z\leq Z_{\text{max}}$. In \cite{2020Liu_CZCP}, it has been shown that $Z_{\text{max}}=N/2$.
\begin{definition}[\cite{2020Liu_CZCP}]
	\label{definition:perfect:CZCP}
	An $(N,Z)$-CZCP is called a perfect $(N,Z)$-CZCP if $Z=\frac{N}{2}$ when $N$ is an even positive integer.
\end{definition} Clearly, the maximum ZCZ width is achieved for a perfect CZCP. Therefore, they are definitely optimal CZCPs.

\subsection{Generalized Boolean Functions (GBFs)}
A $\mathbb{Z}_q$-valued function from $\mathbb{Z}_2^n$ to $\mathbb{Z}_q$ with $n$ variables is said to be a generalized Boolean function (GBF). A GBF  $f: \mathbb{Z}_2^n \rightarrow \mathbb{Z}_q$ can be expressed as a $\mathbb{Z}_q$-linear combination of the $2^n$ monomials $\{1,x_1,\hdots,x_{n},x_1 x_2,\dots, x_1 x_2 x_3,\hdots, x_1 x_2\hdots x_{n}\}$. Note that the product of any $k$ different variables among $x_1,x_2,\cdots, x_{n}$ refers to a monomial of degree $k$ for $n$ variables $x_1,x_2,\cdots,x_{n}\in\mathbb{Z}_2$. For each GBF of $n$-variables, one can associate it with a finite sequence of length $2^n$. Let $f(x_1,x_2,\hdots,x_{n})$ be a GBF and $\mathbf{u}_i \equiv (u_{i,1},u_{i,2},\hdots,u_{i,n})$ be the binary vector 
representation of the index $i$,  where $0\leq i\leq 2^n-1$, i.e., $i = \sum_{l=1}^{n} u_{i,l} 2^{l-1}$. Let $f_i= f(\mathbf{u}_i)= f(u_{i,1},u_{i,2},\hdots,u_{i,n})$ $\forall \ i$.
The $\mathbb{Z}_q$-valued sequence associated with $f$ is denoted by $\textbf{\textit{f}}$ and is defined by $\textbf{\textit{f}}=\big[f_0,f_1,\hdots,f_{2^n-1}\big]$. Clearly, the sequence $\textbf{\textit{f}}$ is a $q$-ary sequence of length $2^n$. The complex-valued sequence with $q$-th roots of unity elements associated with $f$ is denoted by $\psi_q(\textbf{\textit{f}})$  and defined by $\psi_q(\textbf{\textit{f}})= \big[\omega_q^{f_0}, \omega_q^{f_1}, \cdots, \omega_q^{f_{2^n-1}}\big]$. It is worth mentioning that the elements $f_i$'s represent data to be communicated, and the complex-valued sequence $\psi_q(\textbf{\textit{f}})$ represents the complex modulated sequence with $q$-th roots of unity. We will use the following notation throughout the manuscript,
{\small \begin{equation}
\label{notation:rho:ACCF}
\rho^{(q)}_{f,g}(\tau)\equiv \rho_{\psi_q(\textbf{\textit{f}}),\psi_q(\textbf{\textit{g}})}(\tau).
\end{equation}}

The truncated sequence corresponding to the GBF $f$ is denoted by $\textbf{\textit{f}}^{(L)}$ in which the first and the last $L$ elements are discarded from the sequence $\textbf{\textit{f}}$. For example, we consider a binary sequence $\textbf{\textit{f}}=(1100111101)$ of length $N=10$. Then, $\textbf{\textit{f}}^{(3)}=(\underbrace{110}0111\underbrace{101})=(0111)$ of length $N-2L=4$ with $L=3$ and binary elements.

\begin{lemma}[\cite{2021HuangCOML}]
	\label{Huang:Lemma:CZCP}
	Let $\pi$ be a permutation of $\{1,2,\cdots, n-2\}$ for any integer $n\geq 4$. Consider a Boolean function $f$ defined by 
	\begin{align}
	\label{Huang:equ:CZCP}
	&f=\bar{x}_{n}\sum_{k=1}^{n-3}\bar{x}_{\pi(k)}\bar{x}_{\pi(k+1)}+x_n\sum_{k=1}^{n-3}x_{\pi(k)}x_{\pi(k+1)} \nonumber \\
	&\qquad +x_nx_{\pi(1)}+\sum_{k=1}^{n-1}c_kx_k+c_0,
	\end{align} where $\bar{x}=1-x$ and $c_k\in \mathbb{Z}_2$ for $0\leq k \leq n-1$. If $\nu=0$ or $\{\pi(1),\pi(2),\cdots, \pi(\nu)\}=\{1,2,\cdots, \nu\}$ for an integer $\nu \leq n-3$, the sequence pair $(\textbf{\textit{f}}^{(L)}, \left(\textbf{\textit{f}}+\textbf{x}_n)^{(L)}\right)$ forms a binary $(2^{n-1}+2^{\nu+1}, 2^{\pi(\nu+1)-1}+2^{\nu}-1)$-CZCP with $L=2^{n-2}-2^{\nu}$.
\end{lemma}


\section{Proposed Construction of Root Cross $Z$-Complementary Sequences of Length $2^{n-1}+2^{\nu+1}$}
\label{sec:propo:constru}
In this section, we propose a novel construction method for new $q$-ary RCZCPs based on GBF and two roots of unity. 
\subsection{Proposed Construction of RCZCPs}
\begin{algorithm}
	\caption{Proposed Algorithm for New RCZCPs}
	\begin{algorithmic}[1] \label{th:1:rczcp}
		\INPUT Positive integers: $n \ (\geq 4)$, $k_1$ and $k_2$, 
		 permutation:\\ $\pi$, a partition: $P$, $c_i\in \mathbb{Z}_q$ for $0$ $\leq$ $i$ $\leq$ $n-1$, $\nu$, and $L=2^{n-2}-2^{\nu}$ \\
		\OUTPUT A pair of sequences: $\left(\textbf{\textit{f}}_{P}^{(L)}, \textbf{\textit{g}}_{P}^{(L)}\right)$\\
		
		\STATE Consider two roots of unity $\alpha_1=e^{\frac{2\pi \sqrt{-1}}{k_1}}$ and $\alpha_2=e^{\frac{2\pi \sqrt{-1}}{k_2}}$ with $K=\{k_1,k_2\}$.
		\STATE  Take an arbitrary partition $P=\{R_{k_1}, R_{k_2}\}$ of $\{1,$ $2,$ \\ $\cdots,$ $ n\}$ $=$ $R_{k_1} \cup R_{k_2}$. Let $q=LCM\{2,k_1,k_2\}$.
	\STATE  For $n\geq 4$ and a permutation $\pi$ of $\{1,2,\cdots, n-2\}$, we consider a GBF $ f$ defined by
		\begin{align*}
		&f=\bar{x}_{n}\sum_{i=1}^{n-3}\bar{x}_{\pi(i)}\bar{x}_{\pi(i+1)}+x_n\sum_{i=1}^{n-3}x_{\pi(i)}x_{\pi(i+1)} \nonumber \\
		&\qquad +x_nx_{\pi(1)}+\sum_{i=1}^{n-1}c_ix_i+c_0,
		\end{align*} where $\bar{x}=1-x$ and $c_i\in \mathbb{Z}_q$ for $0\leq i \leq n-1$.
	\STATE  Define another GBF $g:\mathbb{Z}_2^n\rightarrow \mathbb{Z}_{q}$ by $g=f+x_{n}$.
	\STATE	 Construct two $q$-ary sequences $\textbf{\textit{f}}_{P}$  and $\textbf{\textit{g}}_{P}$ as follows:
		 \begin{align*}
			&\psi_{q}(\textbf{\textit{f}}_{P})=\omega_q^{\frac{q}{2}f} \alpha_1^{\sum_{i\in R_{k_1}}x_{i}}\alpha_2^{\sum_{j\in R_{k_2}}x_{j}};  \\
			&\psi_{q}(\textbf{\textit{g}}_{P})=\omega_q^{\frac{q}{2}g}\alpha_1^{\sum_{i\in R_{k_1}}x_{i}}\alpha_2^{\sum_{j\in R_{k_2}}x_{j}}. 
			\end{align*}
	\STATE  If $\{\pi(1),\pi(2),\cdots, \pi(\nu)\}$ $=$ $\{1,2,\cdots, \nu\}$ or $\nu=0$ with $\nu\leq n-3$, we consider two truncated sequences $\textbf{\textit{f}}_{P}^{(L)}$ and $\textbf{\textit{g}}_{P}^{(L)}$ for $L=2^{n-2}-2^{\nu}$.
		\STATE  Then, two truncated sequences $\textbf{\textit{f}}_{P}^{(L)}$ and $\textbf{\textit{g}}_{P}^{(L)}$ form a new $q$-ary $(N,Z)$-RCZCP with $N=2^{n-1}+2^{\nu+1}$ and $Z=2^{\pi(\nu+1)-1}+2^{\nu}-1$.
	\end{algorithmic}
\end{algorithm}

\begin{IEEEproof}
	Please see Appendix.
\end{IEEEproof} 
\begin{remark}
	Since the phases of the constructed CZCPs are determined by the given roots of unity $\alpha_1$ and $\alpha_2$, we call the constructed CZCPs as root CZCPs (RCZCPs). We denote it as $(N,Z)$-RCZCP.
\end{remark}

According to the proposed construction method, $Z$ achieves its maximum value when $\pi(\nu+1)=n-2$ and $\nu=n-3$. Thus, $Z_{\text{max}}=N/2=2^{n-2}+2^{\nu}=2^{n-2}+2^{n-3}$ and $Z=2^{\pi(\nu+1)-1}+2^{\nu}-1=2^{n-3}+2^{\nu}-1=2^{n-3}+2^{n-3}-1$. We have $\text{CZC}_{\text{ratio}}=\frac{Z}{Z_{\text{max}}}=\frac{2^{n-3}+2^{n-3}-1}{2^{n-2}+2^{n-3}} \equiv 2/3$. The following example illustrates our proposed \textit{Algorithm \ref{th:1:rczcp}}.
\begin{example}
	\label{example:non:power:two}
	Let $n=5$ and $\nu=1$ with $\pi=(1, 3, 2)$. We have $L=2^{n-2}-2^{\nu}=2^3-2=6$. We consider $\alpha_1=e^{\frac{2\pi\sqrt{-1}}{2}}$ and $\alpha_2=e^{\frac{2\pi\sqrt{-1}}{3}}$ with $k_1=2, k_2=3$ and $P=\{(1,4),(2,3,5)\}$ with $R_2=(1,4)$ and $R_3=(2,3,5)$. Thus, $q=LCM\{2,k_1,k_2\}=LCM\{2,2,3\}=6$.  Let $[c_1,c_2,c_3,c_4]=[2, 3, 0, 5]$ and $c_0=4$. Then, two $6$-ary sequences  $f_{P}$ and $g_{P}$ of length $2^5=32$ are given by
	\begin{align}
	&f_{P}=3\bar{x}_3\bar{x}_5(\bar{x}_1+\bar{x}_2)+3x_3x_5(x_1+x_2)+3x_1x_5+3x_1 \nonumber \\
	& \qquad +5x_2+2(x_3+x_5), \\
	&g_{P}=3\bar{x}_3\bar{x}_5(\bar{x}_1+\bar{x}_2)+3x_3x_5(x_1+x_2)+3x_1x_5+3x_1 \nonumber \\
	&\qquad +5x_2+2x_3+5x_5.
	\end{align} According to \textit{Algorithm \ref{th:1:rczcp}}, for $L=6$, two truncated sequences  $\textbf{\textit{f}}_{P}^{(6)}$ and $\textbf{\textit{g}}_{P}^{(6)}$ of length $N=2^{n-1}+2^{\nu+1}=2^4+2^2=20$ are given by
	{\small
		\begin{align}
		&\psi_{6}(\textbf{\textit{f}}_{P}^{(6)}) =\omega_6^{[1,4,0,0,2,2,2,5,1,4,2,2,1,1,4,1,0,3,2,2]}; \\
		&\psi_{6}(\textbf{\textit{g}}_{P}^{(6)}) =\omega_6^{[1,4,0,0,2,2,2,5,1,4,5,5,4,4,1,4,3,0,5,5]}.
		\end{align}}We give Fig. \ref{fig:AACF:sum:RCZCP} and Fig. \ref{fig:ACCF:sum:RCZCP} to show the absolute values of the sum of AACFs and ACCFs for two sequences $\textbf{\textit{f}}_{P}^{(6)}$ and $\textbf{\textit{g}}_{P}^{(6)}$, respectively. In this case, we have 
	{\small \begin{align*}
		C_1:&\Big(\Big|\rho_{\psi_{6}(\textbf{\textit{f}}_{P}^{(6)})}(\tau)+\rho_{\psi_{6}(\textbf{\textit{g}}_{P}^{(6)})}(\tau)\Big|\Big)_{\tau=0}^{19} =(40, \textbf{O}_{1\times 5}, 8,0,8,\textbf{O}_{1\times 11}) \nonumber \\
		C_2:&\Big(\Big|\rho_{\psi_{6}(\textbf{\textit{f}}_{P}^{(6)}),\psi_{6}(\textbf{\textit{g}}_{P}^{(6)})}(\tau)+\rho_{\psi_{6}(\textbf{\textit{g}}_{P}^{(6)}),\psi_{6}(\textbf{\textit{f}}_{P}^{(6)})}(\tau)\Big|\Big)_{\tau=0}^{19} \nonumber \\
		&=(0,12,0,8,0,4,0,11,0,4, \textbf{O}_{1\times 10}). 
		\end{align*}}Therefore, two sequences $\textbf{\textit{f}}_{P}^{(6)}$ and $\textbf{\textit{g}}_{P}^{(6)}$ form a $6$-ary $(20, 5)$-RCZCP. This constructed $6$-ary $(20, 5)$-RCZCP has not been investigated before. 
	\begin{center}
		\begin{figure}
			\includegraphics[width=0.95\textwidth]{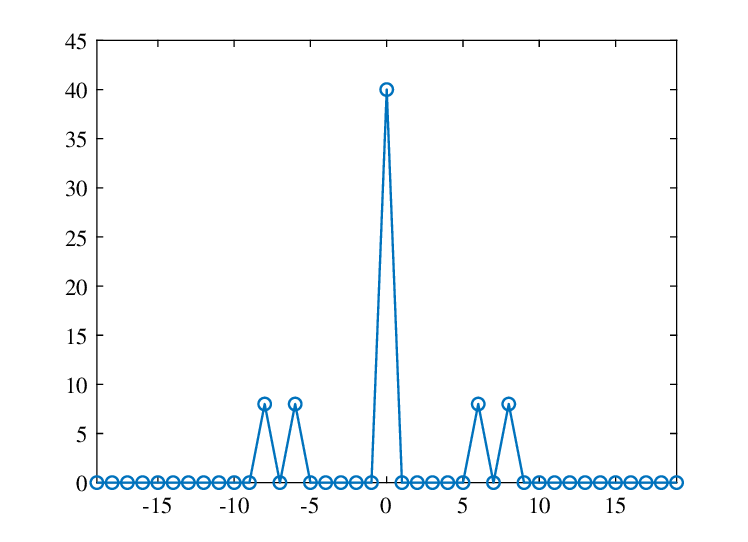}
			\caption{The absolute values of AACF sums of \textit{Example \ref{example:non:power:two}}}
			\label{fig:AACF:sum:RCZCP}
		\end{figure}
	\end{center} 
	\begin{center}
		\begin{figure}
			\includegraphics[width=0.95\textwidth]{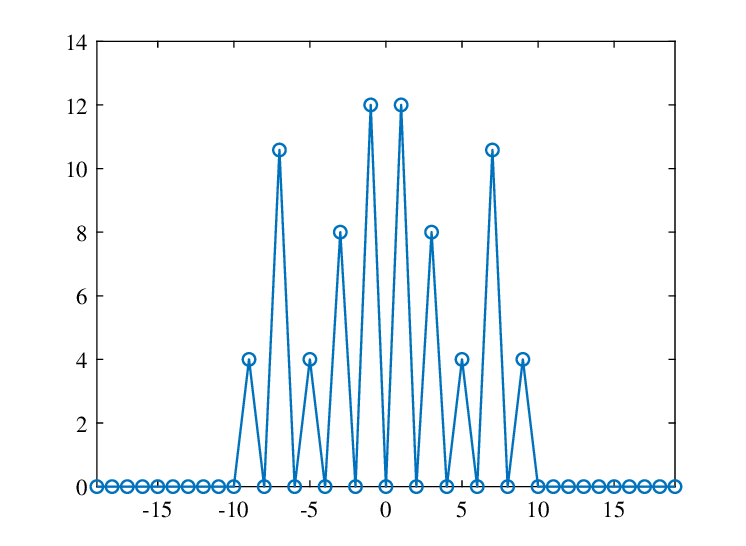}
			\caption{The absolute values of ACCF sums of \textit{Example \ref{example:non:power:two}}}
			\label{fig:ACCF:sum:RCZCP}
		\end{figure}
	\end{center}
\end{example} 

We give the following example by considering only a different partition of $\{1,2,\cdots, n\}$ compared to \textit{Example \ref{example:non:power:two}}.
\begin{example}
	Let $n=5$ and $\nu=1$ with $\pi=(1, 3, 2)$ and $L=2^{n-2}-2^{\nu}=2^3-2=6$. We consider $\alpha_1=e^{\frac{2\pi\sqrt{-1}}{2}}$ and $\alpha_2=e^{\frac{2\pi\sqrt{-1}}{3}}$ with $k_1=2$ and $k_2=3$.  We consider a partition $P=\{(1,3),(2,4,5)\}$ with $R_2=(1,3)$ and $R_3=(2,4,5)$. Thus, $q=LCM\{2,k_1,k_2\}=LCM\{2,2,3\}=6$.  Let $[c_1,c_2,c_3,c_4]=[2, 3, 0, 5]$ and $c_0=4$. Then, two $6$-ary sequences  $f_{P}$ and $g_{P}$ are given by
	\begin{align}
	&f_{P}=3\bar{x}_3\bar{x}_5(\bar{x}_1+\bar{x}_2)+3x_3x_5(x_1+x_2)+3x_1x_5+3x_1 \nonumber \\
	& \qquad +5(x_2+x_4)+3x_3+2x_5, \\
	&g_{P}=3\bar{x}_3\bar{x}_5(\bar{x}_1+\bar{x}_2)+3x_3x_5(x_1+x_2)+3x_1x_5+3x_1 \nonumber \\
	&\qquad +5x_2+3x_3+5(x_4+x_5).
	\end{align} According to \textit{Algorithm \ref{th:1:rczcp}}, two truncated sequences  $\textbf{\textit{f}}_{P}^{(6)}$ and $\textbf{\textit{g}}_{P}^{(6)}$ of length $N=20$ are given by
	{\small
		\begin{align}
		&\psi_{6}(\textbf{\textit{f}}_{P}^{(6)}) =\omega_6^{[2,5,5,5,1,1,2,5,1,4,2,2,1,1,5,2,1,4,1,1]}; \\
		&\psi_{6}(\textbf{\textit{g}}_{P}^{(6)}) =\omega_6^{[2,5,5,5,1,1,2,5,1,4,5,5,4,4,2,5,4,1,4,4]}.
		\end{align}} In this case, we have 
	{\small \begin{align*}
		&C_1:\Big(\Big|\rho_{\psi_{6}(\textbf{\textit{f}}_{P}^{(6)})}(\tau)+\rho_{\psi_{6}(\textbf{\textit{g}}_{P}^{(6)})}(\tau)\Big|\Big)_{\tau=0}^{19} =(40, \textbf{O}_{1\times 5}, 8,0,8,\textbf{O}_{1\times 11}) \nonumber \\
		&C_2:\Big(\Big|\rho_{\psi_{6}(\textbf{\textit{f}}_{P}^{(6)}),\psi_{6}(\textbf{\textit{g}}_{P}^{(6)})}(\tau)+\rho_{\psi_{6}(\textbf{\textit{g}}_{P}^{(6)}),\psi_{6}(\textbf{\textit{f}}_{P}^{(6)})}(\tau)\Big|\Big)_{\tau=0}^{19} \nonumber \\
		&=(0,11,0,14,0,4,0,4,0,4,\textbf{O}_{1\times 10}). 
		\end{align*}}Therefore, two sequences $\textbf{\textit{f}}_{P}^{(6)}$ and $\textbf{\textit{g}}_{P}^{(6)}$ form a new $6$-ary $(20, 5)$-RCZCP. Also, they are different from the RCZCP constructed in \textit{Example \ref{example:non:power:two}}.
\end{example}
\begin{remark}
	Our proposed \textit{Algorithm \ref{th:1:rczcp}} with two different partitions $P=\{R_{k_1}, R_{k_2}\}$ and $P'=\{R_{k'_1}, R_{k'_2}\}$ of $\{1,2,\cdots, n\}$ produces different RCZCPs. 
\end{remark}

\subsection{Comparison with the Existing Works} 
In the literature, there have been four types of construction methods for CZCPs: based on (i) GBFs \cite[Const.1]{2020Liu_CZCP}, \cite[Th.1]{2020AdhikaryTSP}, \cite{2021HuangCOML}, (ii) insertion technique \cite[Th.2]{2020AdhikaryTSP}, (iii) Barker sequences \cite[Th.4]{2020AdhikaryTSP}, and (iv) $Z$-complementary sequences \cite{2020CuilingAdhikaryCOML}, \cite{2021Yang}. 

Based on the proposed \textit{Algorithm \ref{th:1:rczcp}}, the lengths of the constructed $q$-ary RCZCPs are of the form $2^{n-1}+2^{\nu+1}$. Thus, the sequence lengths and the alphabet size of our proposed RCZCPs are flexible compared to the existing works \cite[Th.2]{2020AdhikaryTSP}, \cite[Th.4]{2020AdhikaryTSP}, \cite{2020CuilingAdhikaryCOML} and \cite{2021Yang}. Also, our design strategy has an advantage over \cite[Const.1]{2020Liu_CZCP}, \cite[Th.1]{2020AdhikaryTSP}, and \cite{2021HuangCOML} with regard to the availability of larger class of $q$-ary RCZCPs. For example, $6$-ary CZCPs with sequence lengths $12, 20, 24,\cdots,$ have not been reported before. In addition, we can show that our proposed \textit{Algorithm \ref{th:1:rczcp}} includes \cite[Th. 4]{2021HuangCOML} as a special case.  We take a special case of \textit{Algorithm \ref{th:1:rczcp}} defined by $k_1=1$ and $k_2=1$. Then, we have $q=LCM\{2,1\}=2$. Thus, $\psi_{2}(\textbf{\textit{f}}_{P})$ and $\psi_{2}(\textbf{\textit{g}}_{P})$ become
\begin{align}
\label{cons:seq:alg:2:special}
&\psi_{2}(\textbf{\textit{f}}_{P})=\psi_{2}(\textbf{\textit{f}}); \quad \psi_{2}(\textbf{\textit{g}}_{P})=\psi_{2}(\textbf{\textit{g}}),
\end{align}where the Boolean function $f$ is given by (\ref{Huang:equ:CZCP}) and $g=f+x_n$. Two binary sequences $\textbf{\textit{f}}_{P}^{(L)}$ and  $\textbf{\textit{g}}_{P}^{(L)}$ become $\textbf{\textit{f}}^{(L)}$ and  $\textbf{\textit{g}}^{(L)}$, respectively. Note that the algebraic normal forms of two Boolean functions $f$ and $g$ are same as the algebraic normal forms given in \cite[Th. 4]{2021HuangCOML} (i.e., \textit{Lemma \ref{Huang:Lemma:CZCP}}).

\section{Enumeration for the Constructed RCZCPs of Sequence Lengths $2^{n-1}+2^{\nu+1}$ from \textit{Algorithm \ref{th:1:rczcp}}}
\label{Enume:propo:th1}

By using \textit{Algorithm \ref{th:1:rczcp}}, we obtain $q$-ary RCZCPs with sequence lengths $2^{n-1}+2^{\nu+1}$. We now determine the number of cross $Z$-complementary sequences and pairs constructed by our proposed \textit{Algorithm \ref{th:1:rczcp}}. Based on our proposed construction algorithm, there are $(n-2)!/2$ choices for the permutation $\pi$ defined over $\{1,2,\cdots, n-2\}$. The condition $\{\pi(1),\pi(2),\cdots, \pi(\nu)\}=\{1,2,\cdots, \nu\}$ of $\pi$ implies that there are $(n-\nu-2)!/2$ choices for the permutation $\pi$ and $q^{n}$ choices for $c_0,c_1,c_2,\cdots, c_{n-1}$ in the expression of $f_P$, where $q=LCM\{2,k_1,k_2\}$. Note that our proposed algorithm allows to vary each of $c_0,c_1,c_2,\cdots, c_{n-1}$ in the set $\mathbb{Z}_{q}$. By comparing the algebraic normal forms, two GBFs $f_P$ and $g_P$ are distinct.  In addition, the total number of partitions of $\{1,$ $2,$ $\cdots,$ $ n\}$ into non-empty subsets $\{R_{k_1}, R_{k_2}\}$ is $S(n;2)$, where $S(n;2)$ denotes the Stirling number of the second kind\footnote{The Stirling number of the second kind refers to the total number of ways to partition a finite set of size $n$ into $r$ non-empty subsets. It is usually denoted by $S(n;r)$. For example, $S(4,2)=7$.}  \cite[eq.(1.94a)]{2011Stanley}. Also, our proposed \textit{Algorithm \ref{th:1:rczcp}} with two distinct partitions produces different RCZCPs. Therefore, the total number of distinct sequences constructed by \textit{Algorithm \ref{th:1:rczcp}} is given by
\begin{align}
\label{total:const:seq:pro:alg:2}
N(n,q,P,\nu)&=S(n;2)\cdot \frac{(n-\nu-2)!}{2}\cdot q^n\cdot 2 \nonumber \\
&=S(n;2)\cdot (n-\nu-2)!\cdot q^n.
\end{align}

We now count the total number of distinct sequences constructed from \cite{2021HuangCOML} (i.e., \textit{Lemma \ref{Huang:Lemma:CZCP}}). There are $(n-\nu-2)!/2$ choices for the permutation $\pi$ with $\{\pi(1),\pi(2),\cdots, \pi(\nu)\}=\{1,2,\cdots, \nu\}$ and $2^{n}$ choices for $c_0,c_1,c_2,\cdots, c_{n-1}$. Note that all the sequences $\textbf{\textit{f}}$ and $\textbf{\textit{f}}+\textbf{\textit{x}}_n$ are different by comparing their algebraic normal forms. Therefore, the total number of distinct sequences constructed by \textit{Lemma \ref{Huang:Lemma:CZCP}} is given by 
\begin{align}
\label{total:const:seq:Hunag}
N(n,2,\nu)&=\frac{(n-\nu-2)!}{2}\cdot 2^n\cdot 2 
=(n-\nu-2)!\cdot 2^n.
\end{align} 

According to (\ref{total:const:seq:pro:alg:2}) and (\ref{total:const:seq:Hunag}), we have $N(n,q,P,\nu)\geq N(n,2,\nu)$ as $S(n;2)\geq 1$. Thus, we obtain many new RCZCPs with length $2^{n-1}+2^{\nu+1}$ from \textit{Algorithm \ref{th:1:rczcp}} compared to \cite{2021HuangCOML}. We now count the constructed sequences arising from \cite[Th. 1]{2020AdhikaryTSP}. There are $(n-2)!/2$ choices for the permutation $\pi$, $q^{n-2}$ choices for $e_k$ and $q^{n-2}$ choices for $f_k$ in \cite[Th. 1]{2020AdhikaryTSP}.  Therefore, we can obtain 
\begin{align}
\label{total:const:seq:Avik}
N(n,q)&=\frac{(n-2)!}{2}\cdot q^{2(n-2)}\cdot 2 =(n-2)!\cdot q^{2(n-2)}
\end{align} distinct sequences from \cite[Th. 1]{2020AdhikaryTSP}.
\begin{remark}
	According to (\ref{total:const:seq:pro:alg:2}), (\ref{total:const:seq:Hunag}) and (\ref{total:const:seq:Avik}), for $n=5,\nu=0$ and $q=2$, we can construct $2880$ binary cross $Z$-complementary sequences of length $18$ from our proposed \textit{Algorithm \ref{th:1:rczcp}} as $S(5;2)=15$. In comparison, one can obtain $384$ and $192$ binary sequences of length $18$ from \cite[Th. 1]{2020AdhikaryTSP} and \cite{2021HuangCOML}, respectively. Thus, the proposed \textit{Algorithm \ref{th:1:rczcp}} offers much larger class of RCZCPs compared to existing works.
\end{remark}

We give Table  \ref{table:enumerate:previous:RCZCPs} to compare the existing works \cite{2021HuangCOML},  \cite[Th. 1]{2020AdhikaryTSP} and our proposed \textit{Algorithm \ref{th:1:rczcp}}.
\begin{center}
	\captionof{table}{Comparison with Existing Works \cite{2021HuangCOML}, \cite[Th. 1]{2020AdhikaryTSP} and Our Proposed \textit{Algorithm \ref{th:1:rczcp}} with $n=4$ and $5$. \\ Here, $\surd$ stands for the availability of certain CZCP.}  \label{table:enumerate:previous:RCZCPs}
	\resizebox{\linewidth}{!}{
		\renewcommand{\arraystretch}{1.5}
		\begin{tabular}{|l|l|l|l|l|l|l|l|l|}
			\hline
			\multicolumn{1}{|c|}{$n$} & \multicolumn{1}{c|}{$S(n;2)$} & \multicolumn{1}{c|}{$q$} & \multicolumn{1}{c|}{$\nu$} & \multicolumn{1}{c|}{CZCPs \cite{2021HuangCOML}} & \multicolumn{1}{c|}{CZCPs \cite[Th. 1]{2020AdhikaryTSP}} & \multicolumn{1}{c|}{RCZCPs} & \multicolumn{1}{c|}{Length} & \multicolumn{1}{c|}{$\text{CZC}_{\text{ratio}}$}\\ 
			\hline
			\multicolumn{1}{|c|}{} & \multicolumn{1}{c|}{} & \multicolumn{1}{c|}{2} & \multicolumn{1}{c|}{0} & \multicolumn{1}{c|}{$\surd$} & \multicolumn{1}{c|}{$\surd$} & \multicolumn{1}{c|}{$\surd$} & \multicolumn{1}{c|}{10} & \multicolumn{1}{c|}{$2/5$}\\ 
			\cline{4-9}
			\multicolumn{1}{|c|}{} & \multicolumn{1}{c|}{} & \multicolumn{1}{c|}{} & \multicolumn{1}{c|}{1} & \multicolumn{1}{c|}{$\surd$} & \multicolumn{1}{c|}{-} & \multicolumn{1}{c|}{$\surd$} & \multicolumn{1}{c|}{12} & \multicolumn{1}{c|}{$\equiv 2/3$}\\ 
			\cline{3-9}
			\multicolumn{1}{|c|}{4} & \multicolumn{1}{c|}{7} & \multicolumn{1}{c|}{4} & \multicolumn{1}{c|}{0} & \multicolumn{1}{c|}{-} & \multicolumn{1}{c|}{$\surd$} & \multicolumn{1}{c|}{$\surd$} & \multicolumn{1}{c|}{10} & \multicolumn{1}{c|}{$2/5$}\\ 
			\cline{4-9}
			\multicolumn{1}{|c|}{} & \multicolumn{1}{c|}{} & \multicolumn{1}{c|}{} & \multicolumn{1}{c|}{1} & \multicolumn{1}{c|}{-} & \multicolumn{1}{c|}{-} & \multicolumn{1}{c|}{$\surd$} & \multicolumn{1}{c|}{12} & \multicolumn{1}{c|}{$\equiv 2/3$}\\ 
			\cline{3-9}
			\multicolumn{1}{|c|}{} & \multicolumn{1}{c|}{} & \multicolumn{1}{c|}{6} & \multicolumn{1}{c|}{0} & \multicolumn{1}{c|}{-} & \multicolumn{1}{c|}{$\surd$} & \multicolumn{1}{c|}{$\surd$} & \multicolumn{1}{c|}{10} & \multicolumn{1}{c|}{$2/5$}\\ 
			\cline{4-9}
			\multicolumn{1}{|c|}{} & \multicolumn{1}{c|}{} & \multicolumn{1}{c|}{} & \multicolumn{1}{c|}{1} & \multicolumn{1}{c|}{-} & \multicolumn{1}{c|}{-} & \multicolumn{1}{c|}{$\surd$} & \multicolumn{1}{c|}{12} & \multicolumn{1}{c|}{$\equiv 2/3$} \\ 
			\hline
			\multicolumn{1}{|c|}{} & \multicolumn{1}{c|}{} & \multicolumn{1}{c|}{} & \multicolumn{1}{c|}{0} & \multicolumn{1}{c|}{$\surd$} & \multicolumn{1}{c|}{$\surd$} & \multicolumn{1}{c|}{$\surd$} & \multicolumn{1}{c|}{18} & \multicolumn{1}{c|}{$4/9$}\\ 
			\cline{4-9}
			\multicolumn{1}{|c|}{} & \multicolumn{1}{c|}{} & \multicolumn{1}{c|}{2} & \multicolumn{1}{c|}{1} & \multicolumn{1}{c|}{$\surd$} & \multicolumn{1}{c|}{-} & \multicolumn{1}{c|}{$\surd$} & \multicolumn{1}{c|}{20} & \multicolumn{1}{c|}{$1/2$}\\ 
			\cline{4-9}
			\multicolumn{1}{|c|}{} & \multicolumn{1}{c|}{} & \multicolumn{1}{c|}{} & \multicolumn{1}{c|}{2} & \multicolumn{1}{c|}{$\surd$} & \multicolumn{1}{c|}{-} & \multicolumn{1}{c|}{$\surd$} & \multicolumn{1}{c|}{24} & \multicolumn{1}{c|}{$\equiv 2/3$}\\ 
			\cline{3-9}
			\multicolumn{1}{|c|}{} & \multicolumn{1}{c|}{} & \multicolumn{1}{c|}{} & \multicolumn{1}{c|}{0} & \multicolumn{1}{c|}{-} & \multicolumn{1}{c|}{$\surd$} & \multicolumn{1}{c|}{$\surd$} & \multicolumn{1}{c|}{18} & \multicolumn{1}{c|}{$4/9$}\\ 
			\cline{4-9}
			\multicolumn{1}{|c|}{5} & \multicolumn{1}{c|}{15} & \multicolumn{1}{c|}{4} & \multicolumn{1}{c|}{1} & \multicolumn{1}{c|}{-} & \multicolumn{1}{c|}{-} & \multicolumn{1}{c|}{$\surd$} & \multicolumn{1}{c|}{20} & \multicolumn{1}{c|}{$1/2$}\\ 
			\cline{4-9}
			\multicolumn{1}{|c|}{} & \multicolumn{1}{c|}{} & \multicolumn{1}{c|}{} & \multicolumn{1}{c|}{2} & \multicolumn{1}{c|}{-} & \multicolumn{1}{c|}{-} & \multicolumn{1}{c|}{$\surd$} & \multicolumn{1}{c|}{24} & \multicolumn{1}{c|}{$\equiv 2/3$}\\ 
			\cline{3-9}
			\multicolumn{1}{|c|}{} & \multicolumn{1}{c|}{} & \multicolumn{1}{c|}{} & \multicolumn{1}{c|}{0} & \multicolumn{1}{c|}{-} & \multicolumn{1}{c|}{$\surd$} & \multicolumn{1}{c|}{$\surd$} & \multicolumn{1}{c|}{18} & \multicolumn{1}{c|}{$4/9$}\\ 
			\cline{4-9}
			\multicolumn{1}{|c|}{} & \multicolumn{1}{c|}{} & \multicolumn{1}{c|}{6} & \multicolumn{1}{c|}{1} & \multicolumn{1}{c|}{-} & \multicolumn{1}{c|}{-} & \multicolumn{1}{c|}{$\surd$} & \multicolumn{1}{c|}{20} & \multicolumn{1}{c|}{$1/2$}\\ 
			\cline{4-9}
			\multicolumn{1}{|c|}{} & \multicolumn{1}{c|}{} & \multicolumn{1}{c|}{} & \multicolumn{1}{c|}{2} & \multicolumn{1}{c|}{-} & \multicolumn{1}{c|}{-} & \multicolumn{1}{c|}{$\surd$} & \multicolumn{1}{c|}{24} & \multicolumn{1}{c|}{$\equiv 2/3$}\\ 
			\hline
	\end{tabular}}
\end{center}

\section{Conclusion and Future Research}
\label{sec:con}
In this paper, we have constructed a new class of $q$-ary RCZCPs with both flexible sequence lengths and large $\text{CZC}_{\text{ratio}}$ by utilizing binary CZCPs and two roots of unity. Through enumeration, we have shown that our proposed construction method can produce a larger class of RCZCPs compared to the existing works. An interesting future research would be to construct more RCZCPs of other sequence lengths and large $\text{CZC}_{\text{ratio}}$. 

\section*{Appendix}
\begin{center}
	{(PROOF OF {ALGORITHM \ref{th:1:rczcp}})}
\end{center}

\begin{IEEEproof}
	According to the given $\alpha_1,\alpha_2$, two sequences $f_{P}$ and $g_{P}$ represent two $q$-ary sequences of length $2^n$ given by
	\begin{align}
	&f_{P}=\frac{q}{2}f+\frac{q}{k_1}\sum_{i\in R_{k_1}}x_{i}+\frac{q}{k_2}\sum_{j\in R_{k_2}}x_{j}; \label{cons:seq:th:alg:f:simp} \\
	&g_{P}=\frac{q}{2}g+\frac{q}{k_1}\sum_{i\in R_{k_1}}x_{i}+\frac{q}{k_2}\sum_{j\in R_{k_2}}x_{j}. \label{cons:seq:th:alg:g:simp}
	\end{align}	For given an integer $x$, we consider $y=x+\tau$. Let $(x_1,x_2,\cdots, x_n)$ and  $(y_1,y_2,\cdots, y_n)$ be the binary representations of $x$ and $y$, respectively. The element $f_{P}(x_1,x_2,\cdots, x_n)$ is simply denoted by $f_{P}^{(x)}$ and the element $g_{P}(x_1,x_2,\cdots, x_n)$ is denoted by $g_{P}^{(x)}$. To prove \textit{Algorithm \ref{th:1:rczcp}}, we first introduce the following lemma.
	\begin{lemma}
		\label{lemm:th:1}
		According to (\ref{cons:seq:th:alg:f:simp}) and (\ref{cons:seq:th:alg:g:simp}), we have
		\begin{align}
		\label{identity:alg:2}
		\frac{2}{k_1}\sum_{i\in R_{k_1}}(x_{i}-y_{i})+\frac{2}{k_2}\sum_{j\in R_{k_2}}(x_{j}-y_{j})=c
		\end{align} for some real constant $c$.
	\end{lemma}
	\begin{IEEEproof}
		\textit{Case-I}: $x_{i}= y_{i}$ and $x_{j}= y_{j}$. In this case, we have 
		\begin{align}
		\frac{2}{k_1}\sum_{i\in R_{k_1}}(x_{i}-y_{i})+\frac{2}{k_2}\sum_{j\in R_{k_2}}(x_{j}-y_{j}) =0.
		\end{align}
		\newline \textit{Case-II}: $x_{i}= y_{i}$ and $x_{j}\neq y_{j}$. Since  $x_{j}$ and $y_{j}$ are binary elements, we have 
		\begin{align}
		\frac{2}{k_1}\sum_{i\in R_{k_1}}(x_{i}-y_{i})+\frac{2}{k_2}\sum_{j\in R_{k_2}}(x_{j}-y_{j}) =\frac{2}{k_2}|R_{k_2}|.
		\end{align}
		\newline \textit{Case-III}: $x_{i}\neq y_{i}$ and $x_{j}= y_{j}$. Since  $x_{i}$ and $y_{i}$ are binary elements, we have 
		\begin{align}
		\frac{2}{k_1}\sum_{i\in R_{k_1}}(x_{i}-y_{i})+\frac{2}{k_2}\sum_{j\in R_{k_2}}(x_{j}-y_{j}) =\frac{2}{k_1}|R_{k_1}|.
		\end{align}
		\newline \textit{Case-IV}: $x_{i}\neq y_{i}$ and $x_{j}\neq y_{j}$. Since $x_{i}, y_{i}$ and $x_{j}, y_{j}$ are binary elements, we have 
		{\small	\begin{align*}
			\frac{2}{k_1}\sum_{i\in R_{k_1}}(x_{i}-y_{i})+\frac{2}{k_2}\sum_{j\in R_{k_2}}(x_{j}-y_{j}) =\frac{2}{k_1}|R_{k_1}|+\frac{2}{k_2}|R_{k_2}|.
			\end{align*}} By combining the above four cases, we can see that $\frac{2}{k_1}\sum_{i\in R_{k_1}}(x_{i}-y_{i})+\frac{2}{k_2}\sum_{j\in R_{k_2}}(x_{j}-y_{j})=c$ for some real constant $c$.
	\end{IEEEproof}
	
	By using (\ref{cons:seq:th:alg:f:simp}), (\ref{cons:seq:th:alg:g:simp}) and (\ref{identity:alg:2}), we have 
	\begin{align}
	\label{aacf:sum:cond:root:f:g}
	&\rho_{\psi_{q}(\textbf{\textit{f}}_{P})}(\tau)+\rho_{\psi_{q}(\textbf{\textit{g}}_{P})}(\tau) \nonumber \\ 
	&\qquad \qquad =\sum_{x}\omega_{q}^{f_{P}^{(x)}-f_{P}^{(y)}}+
	\omega_{q}^{g_{P}^{(x)}-g_{P}^{(y)}}  \nonumber \\
	&\qquad \qquad = \sum_{x}\omega_{2}^c\left(\omega_{2}^{f_x-f_y}+
	\omega_{2}^{g_x-g_y}\right)  \nonumber \\
	&\qquad \qquad = \omega_{2}^c\sum_{x}\left(\omega_{2}^{f_x-f_y}+
	\omega_{2}^{g_x-g_y}\right)  \nonumber \\
	&\qquad \qquad =\omega_{2}^c\left(\rho_{\psi_2(\textbf{\textit{f}})}(\tau)+\rho_{\psi_2(\textbf{\textit{g}})}(\tau)\right),
	\end{align}where the element $f(x_1,x_2,\cdots, x_n)$ is denoted by $f_{x}$. According to (\ref{aacf:sum:cond:root:f:g}), we have
	\begin{align}
	\label{C1:aacf:sum:alg:2}
	&\rho_{\psi_{q}(\textbf{\textit{f}}_{P}^{(L)})}(\tau)+\rho_{\psi_{q}(\textbf{\textit{g}}_{P}^{(L)})}(\tau) \nonumber \\ 
	\vspace{0.65cm}
	&\qquad \qquad = \omega_{2}^c\left(\rho_{\psi_2(\textbf{\textit{f}}^{(L)})}(\tau)+\rho_{\psi_2(\textbf{\textit{g}}^{(L)})}(\tau)\right).
	\end{align} Also, we have 
	\begin{align}
	\label{cross:cond:root:f:g:alg:2}
	&\rho_{\psi_{q}(\textbf{\textit{f}}_{P}),\psi_{q}(\textbf{\textit{g}}_{P})}(\tau)+\rho_{\psi_{q}(\textbf{\textit{g}}_{P}),\psi_{q}(\textbf{\textit{f}}_{P})}(\tau) \nonumber \\
	\vspace{0.65cm}
	& =\sum_{x}\omega_{q}^{f_{P}^{(x)}-g_{P}^{(y)}}+
	\omega_{q}^{g_{P}^{(x)}-f_{P}^{(y)}}  \nonumber \\
	&= \sum_{x}\omega_{2}^{\frac{2}{k_1}\sum_{i\in R_{k_1}}(x_{i}-y_{i})+\frac{2}{k_2}\sum_{j\in R_{k_2}}(x_{j}-y_{j})}   \nonumber \\
	& \qquad \qquad \cdot \left(\omega_{2}^{f_x-g_y}+
   \omega_{2}^{g_x-f_y}\right) \nonumber \\
	&= \sum_{x}\omega_{2}^c\left(\omega_{2}^{f_x-g_y}+
	\omega_{2}^{g_x-f_y}\right)  \nonumber \\
	& = \omega_{2}^c\sum_{x}\left(\omega_{2}^{f_x-g_y}+
	\omega_{2}^{g_x-f_y}\right)  \nonumber \\
	&=\omega_{2}^c\left(\rho_{\psi_2(\textbf{\textit{f}}),\psi_2(\textbf{\textit{g}})}(\tau)+\rho_{\psi_2(\textbf{\textit{g}}),\psi_2(\textbf{\textit{f}})}(\tau)\right).
	\end{align} 
	
	Based on (\ref{cross:cond:root:f:g:alg:2}), we have
	\begin{align}
	\label{accf:sum:cross:alg:2}
	&\rho_{\psi_{q}(\textbf{\textit{f}}_{P}^{(L)}),\psi_{q}(\textbf{\textit{g}}_{P}^{(L)})}(\tau)+\rho_{\psi_{q}(\textbf{\textit{g}}_{P}^{(L)}),\psi_{q}(\textbf{\textit{f}}_{P}^{(L)})}(\tau) \nonumber \\
	\vspace{0.7cm}
	&\quad =\omega_{2}^c\left(\rho_{\psi_2(\textbf{\textit{f}}^{(L)}),\psi_2(\textbf{\textit{g}}^{(L)})}(\tau)+\rho_{\psi_2(\textbf{\textit{g}}^{(L)}),\psi_2(\textbf{\textit{f}}^{(L)})}(\tau)\right).
	\end{align} In \textit{Lemma \ref{Huang:Lemma:CZCP}}, for $\{\pi(1),\pi(2),\cdots, \pi(\nu)\}$ $=$ $\{1,2,\cdots, \nu\}$ or $\nu=0$ with $L=2^{n-2}-2^{\nu}$ ($\nu\leq n-3$), it has been proved that two truncated sequences $\textbf{\textit{f}}^{(L)}$ and $\textbf{\textit{g}}^{(L)}$ form a binary $(N,Z)$-CZCP with sequence length $N=2^{n-1}+2^{\nu+1}$ and ZCZ width $Z=2^{\pi(\nu+1)-1}+2^{\nu}-1$. Thus, we have 
	\begin{align}
	&\rho_{\psi_2(\textbf{\textit{f}}^{(L)})}(\tau)+\rho_{\psi_2(\textbf{\textit{g}}^{(L)})}(\tau)=0;\nonumber \\
	&\qquad \qquad  \ \text{for} \ 0<|\tau|\leq Z \ \text{and}\ N-Z\leq |\tau|<N, \label{binary:seed:f:c1}\\
	\vspace{0.7cm}
	&\rho_{\psi_2(\textbf{\textit{f}}^{(L)}),\psi_2(\textbf{\textit{g}}^{(L)})}(\tau)+\rho_{\psi_2(\textbf{\textit{g}}^{(L)}),\psi_2(\textbf{\textit{f}}^{(L)})}(\tau)=0; \nonumber \\
	&\qquad \qquad  \qquad \ \text{for} \ N-Z\leq |\tau|<N. \label{binary:seed:g:c2}
	\end{align} According to (\ref{C1:aacf:sum:alg:2}), (\ref{accf:sum:cross:alg:2}), (\ref{binary:seed:f:c1}) and (\ref{binary:seed:g:c2}), when $L=2^{n-2}-2^{\nu}$, $N=2^{n-1}+2^{\nu+1}$ and $Z=2^{\pi(\nu+1)-1}+2^{\nu}-1$, we now calculate the AACF sum and ACCF sum for the sequences $\psi_{q}(\textbf{\textit{f}}_{P}^{(L)})$ and $\psi_{q}(\textbf{\textit{g}}_{P}^{(L)})$ as follows:
	\begin{align}
	&C_1:\rho_{\psi_{q}(\textbf{\textit{f}}_{P}^{(L)})}(\tau)+\rho_{\psi_{q}(\textbf{\textit{g}}_{P}^{(L)})}(\tau)=0; \nonumber \\
	&\qquad \qquad  \ \text{for} \ 0<|\tau|\leq Z \ \text{and}\ N-Z\leq |\tau|<N, \\
	\vspace{0.7cm}
	&C_2:\rho_{\psi_{q}(\textbf{\textit{f}}_{P}^{(L)}),\psi_{q}(\textbf{\textit{g}}_{P}^{(L)})}(\tau)+\rho_{\psi_{q}(\textbf{\textit{g}}_{P}^{(L)}),\psi_{q}(\textbf{\textit{f}}_{P}^{(L)})}(\tau)=0; \nonumber \\
	&\qquad \qquad \ \qquad \ \text{for} \ N-Z\leq |\tau|<N.
	\end{align}Therefore, two sequences $\psi_{q}(\textbf{\textit{f}}_{P}^{(L)})$ and $\psi_{q}(\textbf{\textit{g}}_{P}^{(L)})$  satisfy both the conditions $C_1$ and $C_2$ given in (\ref{cross:z:com:conditions}) for $L=2^{n-2}-2^{\nu}$.  Consequently, for $L=2^{n-2}-2^{\nu}$, two truncated sequences $\textbf{\textit{f}}_{P}^{(L)}$ and $\textbf{\textit{g}}_{P}^{(L)}$ form a $q$-ary $(N,Z)$-RCZCP with length $N=2^{n-1}+2^{\nu+1}$ and ZCZ width  $Z=2^{\pi(\nu+1)-1}+2^{\nu}-1$. This completes the proof.
\end{IEEEproof} 



\begin{thebibliography}{10}
	\providecommand{\url}[1]{#1}
	\csname url@samestyle\endcsname
	\providecommand{\newblock}{\relax}
	\providecommand{\bibinfo}[2]{#2}
	\providecommand{\BIBentrySTDinterwordspacing}{\spaceskip=0pt\relax}
	\providecommand{\BIBentryALTinterwordstretchfactor}{4}
	\providecommand{\BIBentryALTinterwordspacing}{\spaceskip=\fontdimen2\font plus
		\BIBentryALTinterwordstretchfactor\fontdimen3\font minus
		\fontdimen4\font\relax}
	\providecommand{\BIBforeignlanguage}[2]{{%
			\expandafter\ifx\csname l@#1\endcsname\relax
			\typeout{** WARNING: IEEEtran.bst: No hyphenation pattern has been}%
			\typeout{** loaded for the language `#1'. Using the pattern for}%
			\typeout{** the default language instead.}%
			\else
			\language=\csname l@#1\endcsname
			\fi
			#2}}
	\providecommand{\BIBdecl}{\relax}
	\BIBdecl
	
	\bibitem{2020Liu_CZCP}
	Z.~{Liu}, P.~{Yang}, Y.~L. {Guan}, and P.~{Xiao}, ``Cross {Z}-complementary
	pairs for optimal training in spatial modulation over frequency selective
	channels,'' \emph{IEEE Trans. Signal Process.}, vol.~68, pp. 1529--1543,
	2020.
	
	\bibitem{2011Renzo}
	M.~D. Renzo, H.~Haas, and P.~M. Grant, ``Spatial modulation for
	multiple-antenna wireless systems: a survey,'' \emph{IEEE Commun. Mag.},
	vol.~49, no.~12, pp. 182--191, 2011.
	
	\bibitem{2015YangHanzo}
	P.~Yang, M.~D. Renzo, Y.~Xiao, S.~Li, and L.~Hanzo, ``Design guidelines for
	spatial modulation,'' \emph{IEEE Commun. Surveys Tuts.}, vol.~17, no.~1, pp.
	6--26, 2015.
	
	\bibitem{2020LiuConfISIT}
	Z.~Liu, P.~Yang, Y.~L. Guan, and P.~Xiao, ``Cross {Z}-complementary pairs
	{(CZCPs)} for optimal training in broadband spatial modulation systems,'' in
	\emph{2020 IEEE International Symposium on Information Theory (ISIT)}, 2020,
	pp. 2926--2930.
	
	\bibitem{2020CuilingAdhikaryCOML}
	C.~Fan, D.~Zhang, and A.~R. Adhikary, ``New sets of binary cross
	{Z}-complementary sequence pairs,'' \emph{IEEE Commun. Lett.}, vol.~24,
	no.~8, pp. 1616--1620, 2020.
	
	\newpage
	\bibitem{2020AdhikaryTSP}
	A.~R. {Adhikary}, Z.~{Zhou}, Y.~{Yang}, and P.~{Fan}, ``Constructions of cross
	{Z}-complementary pairs with new lengths,'' \emph{IEEE Trans. Signal
		Process.}, vol.~68, pp. 4700--4712, 2020.
	
	\bibitem{2021HuangCOML}
	Z.~M. {Huang}, C.~Y. {Pai}, and C.~Y. {Chen}, ``Binary cross {Z}-complementary
	pairs with flexible lengths from {B}oolean functions,'' \emph{IEEE Commun.
		Lett.}, vol.~25, no.~4, pp. 1057--1061, 2021.
	
	\bibitem{2021ZhangMesnager}
	\BIBentryALTinterwordspacing
	H.~Zhang, C.~Fan, and S.~Mesnager, ``Constructions of binary cross
	{Z}-complementary pairs with large {CZC} ratio,'' 2021. [Online]. Available:
	\url{https://arxiv.org/abs/2109.04934}
	\BIBentrySTDinterwordspacing
	
	\bibitem{2021Yang}
	M.~Yang, S.~Tian, N.~Li, and A.~R. Adhikary, ``New sets of quadriphase cross
	{Z}-complementary pairs for preamble design in spatial modulation,''
	\emph{IEEE Signal Process. Lett.}, vol.~28, pp. 1240--1244, 2021.
	
	\bibitem{2011Stanley}
	R.~P. Stanley, ``Enumerative combinatorics,'' \emph{Cambridge studies in
		advanced mathematics}, 2011.
	
\end{thebibliography}

\end{document}